\documentclass[12pt,amsmath,amssymb]{iopart}

\usepackage{graphicx}
\usepackage{color}
\usepackage{array}
\eqnobysec
\usepackage[cp1251]{inputenc}
\usepackage[english]{babel}

\begin{document}
 \title{The Schmidt modes of biphoton qutrits: Poincar\'e-sphere representation}

\author{M V Chekhova$^{1,2}$ and M V Fedorov$^{3,4}$}

\address{$^1$Department of Physics, M V Lomonosov Moscow State University, Moscow, Russia\\
$^2$Max-Planck-Institute for the Science of Light, Erlangen, Germany\\
$^3$A M Prokhorov General Physics Institute, Russian Academy of Science, Moscow, Russia.\\
$^4$Moscow Institute of Physics and Technology, Dolgoprudny, Moscow Region, Russia}

\begin{abstract}
For a general-form polarization biphoton qutrit, physically corresponding to a pair of arbitrarily polarized photons in a single frequency and wavevector mode, we explicitly find polarization Schmidt modes. A simple method is suggested for factorizing the state vector and the explicit expressions for the factorizing photon creation operators are found. The degrees of entanglement and polarization of a qutrit are shown to depend directly on the commutation features of the factorizing operators.  Clear graphic representations for the Stokes vectors of the qutrit state as a whole, its Schmidt modes, and factorizing single-photon creation operators are given, based on the Poincar\'e sphere. An experimental scheme is proposed for measuring the parameters of the Schmidt decomposition as well as for demonstrating the operational meaning of qutrit entanglement.
\end{abstract}


\normalsize

\maketitle

\def\thesection{\arabic{section}}


\section{Introduction}
Biphoton qutrits are the simplest two-photon formations finding applications in the science of quantum information. In polarization biphoton qutrits two photons are assumed to have only one degree of freedom, polarization, with identical spatial and spectral features. Such states are usually produced via collinear spontaneous parametric down-conversion (SPDC) degenerate in the frequencies of the photons. The general form of a biphoton qutrit state vector is given by the superposition
\begin{equation}
 \label{QTR}
 |\Psi\rangle=C_1|2_H\rangle+C_2|1_H,\,1_V\rangle+C_3|2_V\rangle,
\end{equation}
where $|2_H\rangle=\frac{a_H^{\dag^2}}{\sqrt{2}}|0\rangle$ and $|2_V\rangle=\frac{a_V^{\dag^2}}{\sqrt{2}}|0\rangle$   are, correspondingly, biphoton state vectors with both photons having either horizontal or vertical polarization, whereas $|1_H,\,1_V\rangle=a_H^\dag a_V^\dag|0\rangle$ is the state vector of two photons with different polarization, horizontal and vertical, $a_H^\dag$ and $a_V^\dag$ are creation operators for photons with horizontal and vertical polarizations. The constants $C_{1,2,3}$ obey the normalization condition $|C_1|^2+|C_2|^2+|C_3|^2=1$. In the general case $C_{1,2,3}$ can be complex. But as the phase of the state $|\Psi\rangle$ as a whole does not affect any measurable characteristics of qutrits, it can be chosen, e.g., in a way making $C_1$ real and positive with the parameters $C_2$ and $C_3$ being complex (if $C_1=0$ any other of two remaining parameters $C_2$ or $C_3$ can be taken real and positive). In the general case this leaves four free parameters determining states of qutrits, e.g., $|C_2|$ and $|C_3|$ plus their phases $\varphi_2$ and $\varphi_3$.

The main two physically important characteristics of biphoton qutrits are their entanglement and polarization. Polarization features of biphoton qutrits have been widely investigated since 1999 \cite{Klyshko-99,ternary,Burl,MCh,orthoexp}. A very interesting and fruitful interpretation of these features is related to the description of biphoton qutrits in terms of the Stokes vectors on the Poincar\'{e} sphere \cite{Burl,MCh}. The qutrit state vectors (\ref{QTR}) were considered as reduced to the form of two creation operators acting on the vacuum state $|\Psi\rangle=NA^\dag B^\dag|0\rangle$. Below, such representation of biphoton qutrits will be referred to as the ``operator factorization$"$, and the operators $A^\dag$ and $B^\dag$, as ``factorizing operators$"$. In \cite{Burl,MCh} the factorizing operators were considered as functions of angles determining the  orientations of the corresponding Stokes vectors on the Poincar\'{e} sphere. The degree of polarization of qutrits was found to be related to the angle between these Stokes vectors. In particular, maximally polarized biphoton qutris were shown to be represented by two parallel Stokes vectors  and unpolarized biphoton qutrits, by a pair of counter-directed Stokes vectors on the Poincar\'{e} sphere.

As to the entanglement, its general definition for arbitrary pure bipartite states can be formulated as the condition that the wave function of the state cannot be factorized as a product of single-particle functions
\begin{equation}
 \label{ent-gen-def}
 \Psi(x_1,x_2)\neq \psi(x_1)\chi(x_2),
\end{equation}
where $x_1$ and $x_2$ are variables of the two particles. This definition is valid for both distinguishable and indistinguishable particles, with either continuous or discrete variables. The definition does not depend on whether the states under consideration can be used for any practical applications (have operational properties) or not. In the case of two indistinguishable particles, their wave functions are obliged to be either symmetric (bosons) or antisymmetric (fermions). If these symmetry requirements give rise to unfactorable wave functions, in accordance with definition (\ref{ent-gen-def}) such states are entangled. The definition (\ref{ent-gen-def}) is strongly supported by the Schmidt-mode analysis. The Schmidt decomposition of the wave function shows the amount of products of the Schmidt modes, the sum of which represents the bipartite wave function. The wave function is factorized and the state is disentangled only if the Schmidt decomposition contains a single term.

In the case of biphoton qutrits its variables are the polarization variables of two photons  $\sigma_1$ and $\sigma _2$, and each of them can take independently one of two values $\sigma_{1,2}=H\,{\rm or}\,V$. The qutrit's wave function is given by a superposition of basic wave functions corresponding to the three terms in the definition of its state vector (\ref{QTR}), $\Psi(\sigma_1,\sigma_2)=C_1\psi_{HH}(\sigma_1,\sigma_2)+C_2\psi_{HV}(\sigma_1,\sigma_2)
+C_3\psi_{VV}(\sigma_1,\sigma_2)$ with the basic wave functions given by
\begin{eqnarray}
 \label{HH}
 \Psi_{HH}(\sigma_1,\sigma_2)=\langle\sigma_1,\sigma_2|2_H\rangle=
 \delta_{\sigma_1,H}\delta_{\sigma_2,H}=\left({1\atop 0}\right)_1\left({1\atop 0}\right)_2,\\
 \nonumber
 \Psi_{HV}(\sigma_1,\sigma_2)=\langle\sigma_1,\sigma_2|1_H,1_V\rangle=\\
 \frac{1}{\sqrt{2}}\Big[\delta_{\sigma_1,H}\delta_{\sigma_2,V}+\delta_{\sigma_1,V}\delta_{\sigma_2,H}\Big]
 \equiv\frac{1}{\sqrt{2}}\left[\left({1\atop 0}\right)_1\otimes\left({0\atop 1}\right)_2
 +\left({0\atop 1}\right)_1\otimes\left({1\atop 0}\right)_2\right],
 \label{wf-HV}\\
 \label{VV}
 \Psi_{HH}(\sigma_1,\sigma_2)=\langle\sigma_1,\sigma_2|2_V\rangle=
 \delta_{\sigma_1,V}\delta_{\sigma_2,V}=\left({0\atop 1}\right)_1\left({0\atop 1}\right)_2.
 \end{eqnarray}
Two forms of writing basis functions via the Kronecker symbols and via two-line columns are equivalent. The first of these two forms is presented to show clearly the meaning of polarization variables and to demonstrate that there is no big difference between the cases of continuous and discrete variables. On the other hand, expressions via columns are more convenient for calculations.

In accordance with the general definition (\ref{ent-gen-def}), the basic states $|2_H\rangle$ and $|2_V\rangle$ are disentangled because their wave functions $\psi_{HH}$ and $\psi_{VV}$ are factorized. In contrast, the state $|1_H,1_V\rangle$ is entangled because its wave function $\psi_{HV}$ (\ref{wf-HV}) is unfactorable.

Note that the last conclusion, about the state $|1_H,1_V\rangle$ being entangled, often meets objections, and the opinions of scientists on this subject are often controversial (for comparison, see, e.g., Refs. \cite{Rubin} and \cite{Lanyon}). In principle, it is impossible to deny that the wave function of this state is given by Eq. (\ref{wf-HV}). Also, it is hardly possible to deny that photons in this state are correlated: if one of the two photons has horizontal polarization, the other one has to be vertically polarized. Thus, the only remaining question concerns the terms: can correlations in the state $|1_H,1_V\rangle$ be interpreted as entanglement or not? Sometimes they are referred to as quantum correlations rather than entanglement \cite{Paskauskas}. We do not think it is reasonable to make such exception for correlations (entanglement) related to the symmetry of wave functions. First, both the definition (\ref{ent-gen-def}) and the Schmidt-mode definition of entanglement do not make any difference between the reasons for which the bipartite wave function is unfactorable, because of the symmetry requirements, or the multimode structure of the states, or additional degrees of freedom of the particles. And second, in multimode biphoton states with more than one degree of freedom the photon-photon correlations (entanglement) are determined by inseparable mixtures of the symmetry-related and multimode entanglement, and it is impossible to separate in this mixture one part to be called quantum correlations and another one to be called entanglement \cite{NJP}. For these reasons we keep the name entanglement for the correlations of photons both in the state $|1_H,1_V\rangle$ and in biphoton qutrits of a general form.

Another way of arguing against calling the state $|1_H,1_V\rangle$ entangled is related to the doubts on whether such states can be used for practical applications in quantum information, and if not, supposedly, they should not be considered as entangled. In particular, according to  \cite{SPK-Mol}, biphoton polarization qutrits cannot be used for teleportation directly but have to be transformed first to states of a higher dimensionality due to the additional degree of freedom, the direction of photon propagation. In Section 7 of this work we consider such a transformation. On the other hand, biphoton qutrits are well known to have practical applications in different fields of quantum information - they are used in protocols of quantum information (see e.g., Refs. \cite{Bechmann, qutr-enc-Kulik}). It is true however that in such protocols the role of the entanglement of qutrit states was not clarified at all. But in this work we suggest a different method of encoding based on the use of the Schmidt decomposition (Section 7). This method will be  shown to be crucially dependent on the degree of entanglement, and it is most efficient in the case of maximally entangled states such as $|1_H,1_V\rangle$. We believe, this is a direct demonstration that the entanglement of biphoton qutrits, including the state $|1_H,1_V\rangle$,  is an operational feature.

Finally, the third objection against the entanglement of states like $|1_H,1_V\rangle =a_H^\dag a_V^\dag|0\rangle$ consists of saying that their state vectors contain the product of the creation operators, and this indicates their factorization and disentanglement. In fact, as known \cite{Burl}, the possibility of representing qutrit state vectors in the form $A^\dag B^\dag |0\rangle$ is a general feature of all biphoton qutrits. But, as we show below in section 3, this ``operator factorization" is accompanied  by the factorization of the qutrit wave function only if $A^\dag=B^\dag$, i.e., only in the case of really disentangled states like $|2_H\rangle$ and $|2_V\rangle$, but not $|1_H,1_V\rangle$. If $A^\dag\neq B^\dag$, the operator factorization exists, whereas the factorization of the wave function does not. So, the operator factorization of the qutrit state vectors cannot be taken as an alternative definition of qutrit entanglement because it contradicts both to the definition (\ref{ent-gen-def}) and to the definition based on the Schmidt-mode analysis.

It is worth emphasizing that in this paper we consider and discuss only pure biphoton polarization states. In the case of mixed two-qubit biphoton polarization states both the definition of entanglement and its quantifiers can be different from those relating to pure states \cite{PRA}. In particular, one of the best measures of entanglement for mixed states is the so-called relative entropy \cite{Vedral}, rather than concurrence or the Schmidt entanglement parameter $K$. And one more comment, we consider here only states of two photons with a single polarization degree of freedom, with both photons belonging to the same spatial and spectral mode. For practical applications the polarization degree of freedom is very special because it provides many possibilities of manipulations. For example, a superposition of single-photon states with the horizontal and vertical polarizations characterizes also an experimentally detectable photon with some intermediate polarization. In contrast, superposition of states of spatially separated photons is not an experimentally observable photon. To emphasize this difference, for pairs of spatially separated photons one can differentiate between the parameters of mode- and particle-entanglement \cite{Wiseman and Vaccaro}. For purely polarization states this is not needed.

In the frame of the definition (\ref{ent-gen-def}) and with the obligatory symmetry of biphoton wave functions taken into account, the Wootters concurrence $C$ \cite{Wootters} characterizing the degree of entanglement of biphoton qutrits of a general form was found in the recent papers  \cite{NJP,JETP} to be given by
\begin{equation}
 \label{conc}
 C=|2C_1C_3-C_2^2|.
\end{equation}
Moreover, it was found that there is a simple and direct relation between the concurrence $C$ and the degree of polarization $P$ for biphoton qutrits,
\begin{equation}
 \label{C_P}
 C^2+P^2=1.
\end{equation}
This relation shows that maximally entangled states ($C=1$) are unpolarized and, oppositely, maximally polarized states ($P=1$) are disentangled ($C=0$) and their wave function is factorized. In particular, as follows from Eqs. (\ref{conc}) and (\ref{C_P}), the state $|1_H,1_V\rangle$ ($C_1=C_3=0,\,C_2=1$) is unpolarized and maximally entangled ($P=0$ and $C=1$). These and other related conclusions are strongly supported by the analysis based on the Schmidt-mode decomposition \cite{NJP,JETP}.

In this work we continue investigating the links between entanglement and polarization of biphoton qutrits, between the approaches based on the Schmidt decomposition and on the analysis of polarization Stokes vectors on the Poincar\'{e} sphere. After a brief reminder about the definitions of the Schmidt modes as the eigenfunctions of the reduced density matrix in the next section, in section 3 we suggest a very simple procedure of the operator factorization. It shows, in particular, that the operator factorization is always doable, and for any given configuration of the qutrit parameters $C_{1.2.3}$ there is only one pair of factorizing operators. In section 4 we describe an alternative procedure of the operator factorization based on the use of the Schmidt-mode representation. By combining these two approaches we establish direct relations between the factorizing operators and operators of photon creation in the Schmidt modes (sections 4 and 5).  In section 6 we describe representations of biphoton qutrits in terms of the Stokes vectors of either the Schmidt modes or the factorizing operators. In this way we establish the full correspondence between the approaches based on the Schmidt-mode and the Stokes-vector analyses. And, finally, in section 7 we describe a scheme of experiment for measuring directly the parameters of the Schmidt decomposition and for its possible application for the information encoding.

\section{The Schmidt modes}
The Schmidt modes are known to be defined as the eigenfunctions of bipartite reduced density matrices.
The latter is defined as the trace over $\sigma_1$ or $\sigma_2$ of the full qutrit's density matrix $\rho=\Psi\otimes\Psi^\dag$. The result is given by \cite{NJP}
\begin{equation}
  \rho_r=Tr_{\sigma_1}\rho=Tr_{\sigma_2}\rho=
  \label{reduced-qutr}
  \left(
 \begin{array}{cc}
  |C_1|^2+\displaystyle\frac{|C_2|^2}{2} &  \displaystyle\frac{C_1C_2^*+C_2C_3^*}{\sqrt{2}} \\
 \displaystyle\frac{C_1^*C_2+C_2^*C_3}{\sqrt{2}} & |C_3|^2+\displaystyle\frac{|C_2|^2}{2}
  \end{array}
  \right).
\end{equation}
 The Schmidt modes $\varphi_\pm$ are the eigenfunctions of $\rho_r$
\begin{equation}
 \label{modes}
 \rho_r\varphi_\pm=\lambda_\pm\varphi_\pm,
 \end{equation}
 where $\lambda_\pm$ are the eigenvalues of $\rho_r$ related to the qutrit concurrence $C$ (\ref{conc}) and degree of polarization $P$ by the equation
\begin{equation}
 \label{lambda}
 \lambda_\pm=\frac{1}{2}\Big(1\pm\sqrt{1-C^2}\Big)=\frac{1\pm P}{2}{,\quad\lambda_++\lambda_-=1} .
\end{equation}
The useful inverse expressions of $C$ and $P$ via $\lambda_\pm$ are
\begin{equation}
 \label{useful}
 C=2\sqrt{\lambda_+\lambda_-},\quad P=\lambda_+-\lambda_-.
\end{equation}

The Schmidt decomposition for the qutrit wave function has the form
\begin{equation}
 \label{Decomp}
 \Psi(\sigma_1,\sigma_2)=\sqrt{\lambda_+}\;\varphi_+(\sigma_1)\varphi_+
 (\sigma_2)+\sqrt{\lambda_-}\;\varphi_-(\sigma_1)\varphi_-(\sigma_2).
\end{equation}
Because of the indistinguishability of the photons, pairs of Schmidt modes ($\varphi_+,\varphi_+$) and ($\varphi_-,\varphi_-$) are formed by identical rather than different functions. This means in fact that each of the two terms in the Schmidt decomposition (\ref{Decomp}) describes a state with two photons in the same mode, $\varphi_+$ or $\varphi_-$. Compared to Eq. (\ref{ent-gen-def}), Eq. (\ref{Decomp}) is more general. It describes both entangled and disentangled states. In terms of the parameters of the Schmidt decomposition biphoton qutrits are disentangled only if $\lambda_+=0$ or $\lambda_-=0$. In these cases the Schmidt decomposition contains only one product of Schmidt modes and the wave function $\Psi(\sigma_1,\sigma_2)$ is factorized. In all other cases the qutrit is entangled, the Schmidt decomposition contains the sum of two products of Schmidt modes, and the wave function $\Psi(\sigma_1,\sigma_2)$ cannot be factorized.

Decomposition of the wave function (\ref{Decomp}) assumes the following Schmidt decomposition for the qutrit state vector:
\begin{equation}
 \label{Decomp-st-vect}
  |\Psi\rangle=\sqrt{\frac{\lambda_+}{2}} a_+^{\dag^{\,2}}|0\rangle+\sqrt{\frac{\lambda_-}{2}} a_-^{\dag^{\,2}}|0\rangle,
\end{equation}
where $a_\pm^\dag$ are photon creation operators for the Schmidt modes $\varphi_\pm$.

The Schmidt decomposition is very useful for presenting in the simplest form such entanglement quantifiers as the Schmidt parameter $K$ and the entropy of the reduced state:
\begin{equation}
 \label{K and entr}
 K=\frac{1}{\sum_\pm\lambda_\pm^2}, \quad S_r=-\sum_\pm\lambda_\pm \log_2\lambda_\pm.
\end{equation}
For two-qubit states the Schmidt parameter $K$ is known \cite{Caves} to be related directly to the concurrence $C$ (\ref{conc}):
\begin{equation}
 \label{C via K}
 C=\sqrt{2(1-K^{-1})}.
\end{equation}

For finding explicitly the Schmidt modes $\varphi_\pm$, one has to solve Eq. (\ref{modes}). In a general form this has never been done. One of the alternative ways to solve this problem is described below in Section 5.

The polarization Stokes vector and the degree of polarization can be defined for the biphoton qutrit  in terms of the reduced density matrix $\rho_r$ as
\begin{equation}
 \label{Stokes}
 {\vec S}=Tr\big(\rho_r{\vec\sigma}\big)\quad{\rm and}\quad P=|{\vec S}|,
\end{equation}
where now ${\vec\sigma}$ is the vector of the Pauli matrices. From here we can deduce immediately that the reduced density matrix of biphoton qutrits (\ref{reduced-qutr}) coincides with the well-known polarization matrix
\begin{equation}
 \label{red-pol}\rho_r=\rho_{pol}=\frac{1}{2}\left(
 \begin{array}{cc}
  1+S_3 &  S_1-iS_2 \\
  S_1+iS_2& 1-S_3
  \end{array}
  \right).
\end{equation}
The polarization matrix is written here in its standard form \cite{LL} although with the components of the Stokes vector re-numbered. Provided that in the collinear SPDC the biphoton beam propagates along the $z$-axis with the horizontal and vertical directions denoted by $x$ and $y$, the Stokes parameter $S_3=1$ corresponds to the linear polarization along the $x$-axis, $S_3=-1$ corresponds to the linear polarization along the $y$-axis, $S_1=\pm 1$ corresponds to the linear polarization along the lines directed at $\pm 45^\circ$ to the $x$-axis in the $xy$-plane, and  $S_2=\pm1$, to circular polarizations.

\section{Operator factorization}

According to the idea of Ref. \cite{Burl}, the qutrit state vector (\ref{QTR}) with arbitrary parameters $C_{1,2,3}$ can be reduced to the form of a product of two new creation operators times the vacuum state,
\begin{equation}
 \label{AB-repr}
 |\Psi\rangle=NA^\dag B^\dag|0\rangle,
\end{equation}
where $N$ is the normalization factor such that $\langle\Psi|\Psi\rangle=1$. The fact that representation (\ref{AB-repr}) always exists and is unique, already mentioned in Ref.~\cite{Burl}, can be very easily proved as follows. Consider an
 operator ${\hat Q}$ such that $|\Psi\rangle={\hat Q}|0\rangle$,
\begin{equation}
 \label{operator Q}
 {\hat Q}=\frac{C_1}{\sqrt{2}}\,a_H^{\dag^2}+C_2\,a_H^\dag a _V^\dag+\frac{C_3}{\sqrt{2}}\,a_V^{\dag^2},
\end{equation}
and a second-order polynomial $Q(x)$ of some variable $x$, associated with ${\hat Q}$,
\begin{equation}
 \label{Polynom}
 Q(x)=\frac{C_1}{\sqrt{2}}\,x^2+C_2\,x+\frac{C_3}{\sqrt{2}}.
\end{equation}
By finding the roots of the equation $Q(x)=0$,
\begin{equation}
 \label{x+-}
 x_{A,B}=\frac{-C_2\pm\sqrt{C_2^2-2C_1C_3}}{\sqrt{2}\,C_1},
\end{equation}
we represent $Q(x)$ as a product of two linear functions of $x$,
\begin{equation}
 \label{product}
 Q(x)=\frac{C_1}{\sqrt{2}}\,(x-x_A)(x-x_B).
\end{equation}
As the operators $a_H^\dag$ and $a_V^\dag$ commute with each other, they can be arbitrarily transposed and combined as simple numbers. Hence, we can immediately write down the operator analog of the algebraic equality (\ref{product}),
\begin{equation}
 \label{Q-factorized}
 {\hat Q}=\frac{C_1}{\sqrt{2}}\,(a_H^\dag-x_A\,a_V^\dag)(a_H^\dag-x_Ba_V^\dag).
\end{equation}
This equation determines the factorizing operators $A^\dag$ and $B^\dag$ of Eq. (\ref{AB-repr}), though it leaves undefined their phases $\varphi_0$ and $-\varphi_0$:
{\begin{equation}
 \label{AB-x+-}
 A^\dag=e^{i\varphi_0}\frac{a_H^\dag-x_A\,a_V^\dag}{\sqrt{1+|x_A|^2}}\quad{\rm and}\quad B^\dag=e^{-i\varphi_0}\frac{a_H^\dag-x_Ba_V^\dag}{\sqrt{1+|x_B|^2}},
\end{equation}}
The phase factors in these definitions do not affect usual commutation rules for the photon annihilation and creation operators in a given mode,
\begin{equation}
 \label{comm-rules}
 [A,A^\dag]\equiv AA^\dag-A^\dag A=1$ and $[B,B^\dag]=1.
 \end{equation}
But they do affect the cross commutators $[A,B^\dag]$ and $[B,A^\dag]=[A,B^\dag]^\dag$. As shown in the following section, these commutators are always real, whereas the parameters $x_{A,B}$ (\ref{x+-}) can be complex, $x_{A,B}=|x_{A,B}|\exp(i\varphi_{A,B})$, where $\varphi_A$ and $\varphi_B$ are phases of $x_A$ and $x_B$. Thus, the phase $\varphi_0$ can be found just from the condition that the cross-commutators are real
\begin{equation}
 \nonumber
 [A,B^\dag]=\frac{e^{-2i\varphi_0}(1+x_A^*x_B)}{\sqrt{(1+|x_A|^2)(1+|x_B|^2)}}=\frac{|1+x_A^*x_B|}{\sqrt{(1+|x_A|^2)(1+|x_B|^2)}}.
 \label{comm-x+-}
\end{equation}
Explicitly, the phase $\varphi_0$ is given by
\begin{equation}
 \label{phi-0}
 \varphi_0=\frac{1}{2}\tan^{-1}
 \left[\frac{|x_A||x_B|\sin(\varphi_B-\varphi_A)}{1+|x_A||x_B|\cos(\varphi_B-\varphi_A)}\right].
\end{equation}
The commutator (\ref{comm-x+-}) is a very important characteristics of the biphoton qutrit states and, in particular, it determines the normalizing factor $N$ in Eq. (\ref{AB-repr})
\begin{equation}
\label{norm-factor}
 N=\frac{1}{\sqrt{1+[A,B^\dag]^2}}.
\end{equation}

\section{Operator factorization in the Schmidt representation}

The Schmidt decomposition (\ref{Decomp}) provides an alternative way for the operator factorization of qutrit state vectors. In this approach we can further specify the earlier introduced parameters, such as the commutator of the operators $A$ and $B^\dag$ (\ref{comm-x+-}), as well as the definition (\ref{AB-x+-}) of the operators $A^\dag$ and $B^\dag$ themselves.

In accordance with Eq. (\ref{Decomp-st-vect}) and (\ref{operator Q}), the operator $\hat{Q}$ in the qutrit defined as $|\Psi\rangle=\hat{Q}|0\rangle$ in the Schmidt-mode representation takes the form \begin{equation}
 \label{Q-Schmidt}
  \hat{Q}=\frac{\sqrt{\lambda_+}\,a_+^{\dag^{\,2}}+\sqrt{\lambda_-}\,a_-^{\dag^{\,2}}}{2}.
\end{equation}
This operator is factorized easily to give  ${\hat Q}=NA^\dag B^\dag$ with the factorizing operators $A^\dag$ and $B^\dag$ given by
\begin{equation}
 \label{AB-SCHM}
 \setlength{\extrarowheight}{0.4cm}
 \begin{array}{c}
 A^\dag=\displaystyle\frac{\lambda_+^{1/4}a_+^\dag
 +i\lambda_-^{1/4}a_-^\dag}{\left(\sqrt{\lambda_+}+\sqrt{\lambda_-}\,\right)^{1/2}}
 \equiv\displaystyle\frac{\lambda_+^{1/4}a_+^\dag+i\lambda_-^{1/4}a_-^\dag}{(1+C)^{1/4}},\\
 B^\dag=\displaystyle\frac{\lambda_+^{1/4}a_+^\dag
 -i\lambda_-^{1/4}a_-^\dag}{\left(\sqrt{\lambda_+}+\sqrt{\lambda_-}\,\right)^{1/2}}
 \equiv\displaystyle\frac{\lambda_+^{1/4}a_+^\dag-i\lambda_-^{1/4}a_-^\dag}{(1+C)^{1/4}}.
 \end{array}
\end{equation}

The commutator of the operators $A$ and $B^\dag$ can be expressed now in terms of the Schmidt eigenvalues $\lambda_{\pm}$, or the concurrence $C$, or the degree of polarization $P$,
\begin{equation}
 \label{Comm-Schm}
 [A,B^\dag]=[B,A^\dag]=\displaystyle\frac{\sqrt{\lambda_+}-\sqrt{\lambda_-}}{\sqrt{\lambda_+}+\sqrt{\lambda_-}}
 =\displaystyle\sqrt{\frac{1-C}{1+C}}=\frac{P}{1+\sqrt{1-P^2}}\,,
\end{equation}
and similarly for the normalization factor in Eq. (\ref{AB-repr}),
\begin{equation}
 \label{Norm-Schm}
  N=\displaystyle\frac{1}{\sqrt{1+[A,B^\dag]^2}}=\frac{\sqrt{\lambda_+}+\sqrt{\lambda_-}}{\sqrt{2}}=
  \sqrt{\frac{1+C}{2}}.
\end{equation}
Eq. (\ref{Comm-Schm}) shows that in the maximally entangled unpolarized states  ($C=1,\,P=0$) the commutator $[A,B^\dag]$ equals zero, which means that the operators $A^\dag$ and $B^\dag$ represent orthogonal modes, as, e.g., $a_H^\dag$ and $a_V^\dag$. In the case of disentangled maximally polarized states ($C=0,\,P=1$) the commutator $[A,B^\dag]$ equals 1, which indicates that the operators $A^\dag$ and $B^\dag$ coincide with each other. Eq. (\ref{Comm-Schm}) can be inverted to express the concurrence $C$ via the commutator $[A,B^\dag]$,
\begin{equation}
 \label{C via comm}
 C=\frac{1-[A,B^\dag]^2}{1+[A,B^\dag]^2}.
\end{equation}
In principle, this formula can be considered as a definition of the concurrence alternative to that of Eq. (\ref{conc}). Though it is hardly easy to find the commutator $[A,B^\dag]$ before finding the concurrence, Eq. (\ref{C via comm}) is important because it shows that for getting disentangled states of biphoton qutrits it is not sufficient to provide the operator factorization. It is necessary to have the commutator equal unity, $[A,B^\dag]=1$. In other words, the operator factorization by no means leads automatically to the factorization of the biphoton wave function and, hence, the operator factorization does not guarantee disentanglement. The simplest and most often discussed example is the state $|1_H,1_V\rangle$. The operator factorization is present automatically with the factorizing operators $A^\dag=a_H^\dag$ and $B^\dag=a_V^\dag$. But the commutator $[A,B^\dag]=[a_H,a_V^\dag]$ equals zero, and hence, there is no factorization in the biphoton wave function (\ref{wf-HV}), the state is maximally entangled and unpolarized.


\section{Finding the Schmidt modes without solving the eigenvalue-eigenfunction equation (\ref{modes})}

Eqs. (\ref{AB-SCHM}) can be inverted to express the Schmidt-mode creation operators $a_+^\dag$ and $a_-^\dag$ in terms of the factorizing operators $A^\dag$ and $B^\dag$
\begin{equation}
 \label{a+,a- via AB}
\setlength{\extrarowheight}{0.2cm}
 \begin{array}{c}
 a_+^\dag =\displaystyle\frac{1}{2}\left(\frac{1+C}{\lambda_+}\right)^{1/4}\big(A^\dag+B^\dag\big),\\
 a_-^\dag=\displaystyle \frac{i}{2}\,\left(\frac{1+C}{\lambda_-}\right)^{1/4}\big(B^\dag-A^\dag\big).
 \end{array}
\end{equation}
By substituting here expressions (\ref{AB-x+-}) for $A^\dag$ and $B^\dag$ via $a_H^\dag$ and $a_V^\dag$, we finally get the general expressions for the Schmidt-mode creation operators,
\begin{eqnarray}
 \nonumber
 a_+^\dag=\displaystyle\frac{1}{2}\left(\frac{1+C}{\lambda_+}\right)^{1/4}
 \left\{\left[\frac{e^{i\varphi_0}}{\sqrt{1+|x_A|^2}}+\frac{e^{-i\varphi_0}}{\sqrt{1+|x_B|^2}}\right]\,a_H^\dag\right.\\
 \label{a+}
 \left.
 -\left[\frac{e^{i\varphi_0}x_A}{\sqrt{1+|x_A|^2}}+\frac{e^{-i\varphi_0}x_B}{\sqrt{1+|x_B|^2}}\right]\,a_V^\dag\right\}
\end{eqnarray}
and
 \begin{eqnarray}
 \nonumber
 a_-^\dag=\displaystyle\frac{1}{2}\left(\frac{1+C}{\lambda_-}\right)^{1/4}
 \left\{\left[\frac{e^{-i\varphi_0}}{\sqrt{1+|x_B|^2}}-\frac{e^{i\varphi_0}}{\sqrt{1+|x_A|^2}}\right]\,a_H^\dag\right.\\
 \label{a-}
  -\left.\left[\frac{e^{-i\varphi_0}x_B}{\sqrt{1+|x_B|^2}}-\frac{e^{i\varphi_0}x_A}{\sqrt{1+|x_A|^2}}\right]\,a_V^\dag\right\}.
\end{eqnarray}
These operators can be checked to obey the usual commutation rules for the creation and annihilation operators in two orthogonal modes,
\begin{equation}
 \label{comm a+ a-}
 [a_+,a_+^\dag]=1, \; [a_-,a_-^\dag]=1,\;[a_+,a_-^\dag]=[a_-,a_+^\dag]=0.
\end{equation}
The Schmidt modes corresponding to the operators $a_+^\dag$ (\ref{a+}) and $a_-^\dag$ (\ref{a-}) are given by
\begin{equation}
 \label{modes via a+-}
 \varphi_\pm(\sigma)=\langle\sigma|a_\pm^\dag|0\rangle\equiv\left({\langle 1_H|a_\pm^\dag|0\rangle}\atop{\langle 1_V|a_\pm^\dag|0\rangle}\right).
\end{equation}

It should be stressed that the Schmidt modes are found here with the help of simple algebraic operations and without solving the eigenvalue-eigenfunction equation (\ref{modes}), which is rather unexpected.

As an example, let us consider a simplified biphoton qutrit of the form
\begin{equation}
 \label{the simplest}
 |\Psi\rangle=N\,a_H^\dag\,\big(\cos\alpha\,a_H^\dag+\sin\alpha\,a_V^\dag\big)|0\rangle.
\end{equation}
This is the case when the qutrit state vector ``automatically$"$ has the form $NA^\dag B^\dag|0\rangle$ and no factorization procedure is needed. The factorizing operators $A^\dag$ and $B^\dag$ are then given by
\begin{equation}
 \label{AB-Example}
 A^\dag=a_H^\dag\quad{\rm and}\quad B^\dag=\cos\alpha\,a_H^\dag+\sin\alpha\,a_V^\dag.
 \end{equation}
Their commutator and normalization constant $N$ are equal to
\begin{equation}
\label{comm-norm-example}
 [A,B^\dag]=\cos\alpha\quad{\rm and}\quad N=\frac{1}{\sqrt{1+\cos^2\alpha}}.
\end{equation}
In the original form (\ref{QTR}), the parameters of the qutrit (\ref{the simplest}) are equal to
\begin{equation}
 \label{C123-example}
 C_1=\frac{\sqrt{2}\;\cos\alpha}{\sqrt{1+\cos^2\alpha}},\quad C_2=\frac{\sin\alpha}{\sqrt{1+\cos^2\alpha}},\quad C_3=0.
\end{equation}
These parameters correspond to the following concurrence $C$, degree of polarization $P$, and the eigenvalues of the reduced density matrix $\lambda_\pm$:
\begin{equation}
 \label{C-P-example}
 C=C_2^2=\frac{\sin^2\alpha}{1+\cos^2\alpha}, \quad P=\sqrt{1-C^2}=\frac{2\cos\alpha}{1+\cos^2\alpha},
\end{equation}

\begin{equation}
 \label{lambda-example}
 \lambda_\pm=\frac{1}2\left(1\pm\sqrt{1-C^2}\right)=\frac{1\pm P}{2}=\frac{1}{2}\,\frac{(1\pm\cos\alpha)^2}{1+\cos^2\alpha}.
\end{equation}
As the factorizing operators (\ref{AB-Example}) of the qutrit (\ref{the simplest}) are known ``automatically$"$, the Schmidt-mode creation operators can be found directly from Eqs. (\ref{a+,a- via AB}) rather than from more complicated equations (\ref{a+}) and (\ref{a-}),
\begin{equation}
 \label{operators-example}
 \setlength{\extrarowheight}{0.3cm}
 \begin{array}{c}
 a_+^\dag=\displaystyle\frac{1}{\sqrt{2(1+\cos\alpha)}}\left[(1+\cos\alpha)\,a_H^\dag+\sin\alpha\,a_V^\dag\right],\\
 a_-^\dag=\displaystyle\frac{i}{\sqrt{2(1-\cos\alpha)}}\left[(-1+\cos\alpha)\,a_H^\dag+\sin\alpha\,a_V^\dag\right].
 \end{array}
\end{equation}
From these equations and with the help of Eqs. (\ref{modes via a+-}) we easily find explicit expressions for the Schmidt modes of the biphoton qutrit (\ref{the simplest})
\begin{eqnarray}
 \label{modes-example+}
 \varphi_+=\frac{1}{\sqrt{2(1+\cos\alpha)}}\left(\begin{array}{c}1+\cos\alpha\\ \sin\alpha\end{array}\right)
 =\left(\begin{array}{c}\cos(\alpha/2)\\ \sin(\alpha/2)\end{array}\right),\\
 \label{modes-example-}
 \varphi_-=\frac{i}{\sqrt{2(1-\cos\alpha)}}\left(\begin{array}{c}-1+\cos\alpha\\\sin\alpha\end{array}\right)
 =i\left(\begin{array}{c}-\sin(\alpha/2)\\ \cos(\alpha/2)\end{array}\right).
\end{eqnarray}

\section{Poincar\'{e}-sphere representation}

\subsection{The Stokes vectors of single-photon states}

Following earlier works \cite{Burl,MCh} let us consider representation of the above-derived results in the space of Stokes vectors on the Poincar\'{e}-sphere. Let us start with the general expressions (\ref{AB-x+-}) for the factorizing operators $A^\dag$ and $B^\dag$. These operators generate a pair of single-photon states and the corresponding wave functions:
 \begin{eqnarray}
 |1_A\rangle=A^\dag|0\rangle\equiv\frac{1}{\sqrt{1+|x_A|^2}}(a_H^\dag-x_Aa_V^\dag)|0\rangle,\\
 |1_B\rangle=B^\dag|0\rangle\equiv\frac{1}{\sqrt{1+|x_B|^2}}(a_H^\dag-x_Ba_V^\dag)|0\rangle
 \end{eqnarray}
and
\begin{equation}
 \label{1A-1B-wf}
 \psi_A=\frac{1}{\sqrt{1+|x_A|^2}}\left(1\atop{-x_A}\right),\;
 \psi_B=\frac{1}{\sqrt{1+|x_B|^2}}\left(1\atop{-x_B}\right),
\end{equation}
where, as shown above, the parameters $x_{A,B}$ can be expressed via the original qutrit parameters by equations (\ref{x+-}). We have dropped the phase factors occurring in operators $A^\dag$ and $B^\dag$ (3.7) because they do not affect the polarization matrices of the states considered below.

\begin{figure}[h]
\centering\includegraphics[width=9cm]{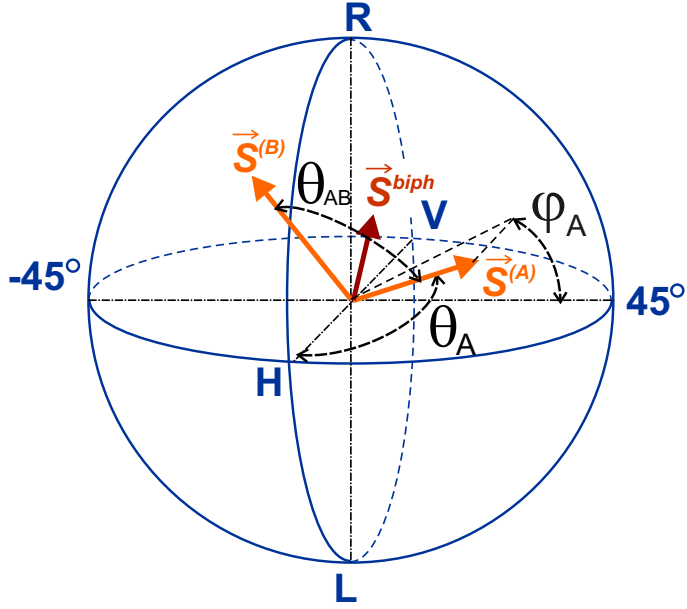}
\caption{{\protect\footnotesize {The Stokes vectors ${\vec S}^{(A)}$ (with spherical coordinates $\theta_A,\varphi_A$) and ${\vec S}^{(B)}$ (\ref{S-A},\ref{S-B}), as well as the biphoton Stokes vector  ${\vec S}^{\,biph}$ (\ref{bf-stokes.v.-comp}), (\ref{Vect S-bph}), (\ref{Sbf=S++S-}), on the Poincar\'{e} sphere. }}}\label{Fig.1}
\end{figure}

The density matrices of the single-photon states $|1_A\rangle$ and $|1_B\rangle$ are easily found from their wave functions to be given by
\begin{equation}
 \label{1A-1B-dens.matr.}
 \rho_A=\frac{1}{1+|x_A|^2}\left(\begin{array}{cc}1 & -x_A^*\\-x_A & |x_A|^2\end{array}\right),
\end{equation}
and the same for $\rho_B$ with the substitution $x_A\rightarrow x_B$. By identifying the density matrices of the states $|1_A\rangle$ and $|1_B\rangle$ with the corresponding polarization matrices $\rho_{pol}^{(A)}$ and $\rho_{pol}^{(B)}$ (having the same form as the polarization matrix in Eq. (\ref{red-pol})) we can find the components of the Stokes vectors ${\vec S}^{\,(A)}$ and ${\vec S}^{\,(B)}$ of the states $|1_A\rangle$ and $|1_B\rangle$:
\begin{equation}
 \label{S-A}
 \setlength{\extrarowheight}{0.1cm}
 \begin{array}{c}
 S_3^{(A)}=(1-|x_A|^2)/(1+|x_A|^2)\equiv\cos\theta_A,\\
  S_1^{(A)}=-2\hbox{Re}(x_A)/(1+|x_A|^2)\equiv\sin\theta_A\,\cos\varphi_A,\\
  S_2^{(A)}=-2\hbox{Im}(x_A)/(1+|x_A|^2)\equiv\sin\theta_A\,\sin\varphi_A,
 \end{array}
\end{equation}
\begin{equation}
 \label{S-B}
 \setlength{\extrarowheight}{0.1cm}
 \begin{array}{c}
  S_3^{(B)}=(1-|x_B|^2)/(1+|x_B|^2)\equiv\cos\theta_B,\\
  S_1^{(B)}=-2\hbox{Re}(x_B)/(1+|x_B|^2)\equiv\sin\theta_B\,\cos\varphi_B,\\
  S_2^{(B)}=-2\hbox{Im}(x_B)/(1+|x_B|^2)\equiv\sin\theta_B\,\sin\varphi_B,
\end{array}
\end{equation}
where $\theta_A$ and $\theta_B$ are the angles between the vectors ${\vec S}^{\,(A)}$ and ${\vec S}^{\,(B)}$ and the $H$-axis on the Poincar\'{e} sphere (Fig.\ref{Fig.1}).
The values $\varphi_A$ and $\varphi_B$ are the angles between the horizontal axis $O$$45^\circ\perp OH$ on the Poincar\'{e} sphere and projections of the Stokes vectors ${\vec S}^{\,(A)}$ and ${\vec S}^{\,(B)}$ on the plane perpendicular to $OH$.

As it follows from Eqs. (\ref{S-A}) and (\ref{S-B}), the absolute values of the Stokes vectors ${\vec S}^{\,(A)}$ and ${\vec S}^{\,(B)}$ are equal unity, $\left|{\vec S}^{\,(A)}\right|=\left|{\vec S}^{\,(B)}\right|=1$. The angle between these vectors $\theta_{AB}$ is determined by their scalar product,
\begin{equation}
 \label{theta-AB}
 \cos\theta_{AB}={\vec S}^{\,(A)}\cdot{\vec S}^{\,(B)}=\cos\theta_A\cos\theta_B+\sin\theta_A\sin\theta_B\cos(\varphi_A-\varphi_B).
\end{equation}

\subsection{Schmidt-mode representation and Poincar\'{e}-sphere images of biphoton qutrits}

\begin{figure}[h]
\centering\includegraphics[width=15cm]{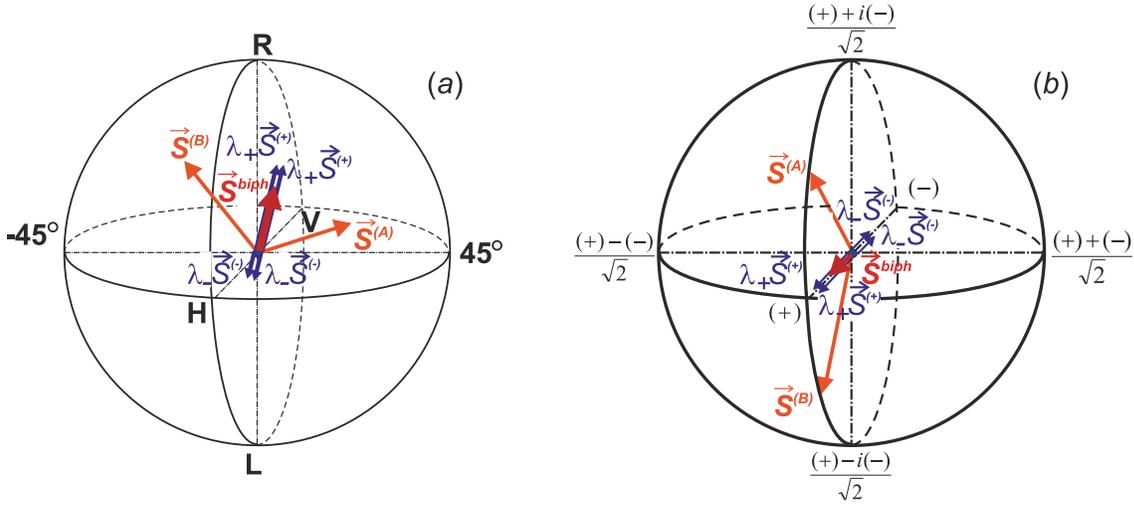}
\caption{{\protect\footnotesize $a$: The Stokes vectors corresponding to the biphoton, to the factoring operators, and to the Schmidt modes multiplied by their eigenvalues, on the Poincar\'{e} sphere; $b$: The same, in the Schmidt-mode representation (with the sphere rotated)}}\label{Fig.2}
\end{figure}

Let us analyze now the Poincar\'{e}-sphere image of the general-form biphton qutrit (\ref{AB-repr}).
One possibility of its investigation consists of finding from this equation the original qutrit's parameters $C_{1,2,3}$,
substituting them into the general expression for the reduced density matrix (\ref{red-pol}), and finding from this last equation
the biphoton Stokes parameters, which involves rather cumbersome transformations and formulas. An alternative, more elegant and even
more informative approach is based on the use of the Schmidt-mode representation. In this representation the Schmidt modes and their single-photon density matrices are very simple:
\begin{equation}
 \label{Schm-single-w.f.}
  \varphi_+=\left(1\atop 0\right),\quad  \varphi_-=\left(0\atop 1\right);\quad \rho_+=\left(\begin{array}{cc}1&0\\0&0\end{array}\right),\quad
  \rho_-=\left(\begin{array}{cc}0&0\\0&1\end{array}\right).
\end{equation}
The Stokes vectors ${\vec S}_+$ and ${\vec S}_-$ corresponding to the Schmidt modes (\ref{Schm-single-w.f.}) have components
\begin{equation}
 \label{1ph-Schm-mode-st.vect}
 S_3^{(+)}=1,\,S_2^{(+)}=S_1^{(+)}=0\quad{\rm and}\quad S_3^{(-)}=-1,\,S_2^{(-)}=S_1^{(-)}=0.
\end{equation}
In terms of the Poincar\'{e} sphere this means that the transition to the Schmidt-mode representation is realized by means of a rotation of the Poincar\'{e} sphere that makes the Stokes vectors of the Schmidt modes directed along the former $(H,V)$ axis. Then this axis turns into the $(+,-)$ axis, where $(+)$ and $(-)$ are the contractions of $\varphi_+$ and $\varphi_-$. The Stokes vectors ${\vec S}^{(+)}$ and ${\vec S}^{(-)}$ point, correspondingly, at the positive and negative directions along this axis (Fig. \ref{Fig.2}).
The biphoton reduced density matrix (\ref{red-pol}) is also very simple in the Schmidt-mode representation
\begin{equation}
 \label{rho-r-Schm}
 \rho_r=\left(\begin{array}{cc}\lambda_+&0\\0&\lambda_-\end{array}\right).
\end{equation}
As well as the one-photon Schmidt-mode Stokes vectors ${\vec S}^{(\,\pm)}$, the biphoton Stokes vector ${\vec S}^{\,biph}$ has only one non-zero component,
\begin{equation}
 \label{bf-stokes.v.-comp}
 S_3^{\,biph}=\lambda_+-\lambda_-=P\quad{\rm with}\quad S_2^{\,biph}=S_1^{\,biph}=0.
\end{equation}
On the Poincar\'{e} sphere the vector ${\vec S}^{\,biph}$ is directed along the same axis $(+,-)$ as the Schmidt-mode Stokes vectors ${\vec S}^{(\,\pm)}$. As usual, the absolute value of ${\vec S}^{\,biph}$ coincides with the degree of polarization per photon $P$ for biphoton polarization qutrits \cite{Burl,MCh},
\begin{equation}
 \label{S3=P}
 \left|{\vec S}^{\,biph}\right|=S_3^{\,biph}=P.
\end{equation}

Finally, the factorizing operators $A^\dag$ and $B^\dag$ are given by superpositions of the Schmidt-mode creation operators (\ref{AB-SCHM}). In the Schmidt-mode representation the one-photon states generated by these operators, $A^\dag|0\rangle$ and$B^\dag|0\rangle$, have the following wave functions
\begin{equation}
 \label{wf-AB-Schm-repr}
 \psi_A=\frac{1}{(1+C)^{1/4}}\left(\begin{array}{c}\lambda_+^{1/4}\\i\lambda_-^{1/4}\end{array}\right),\quad
 \psi_B=\frac{1}{(1+C)^{1/4}}\left(\begin{array}{c}\lambda_+^{1/4}\\-i\lambda_-^{1/4}\end{array}\right)
\end{equation}
and density matrices
\begin{equation}
 \label{d.m.AB Schm}
 \rho_{A,B}=\frac{1}{\sqrt{1+C}}\left(\setlength{\extrarowheight}{0.2cm}
 \begin{array}{cc}\sqrt{\lambda_+}&\mp i(\lambda_+\lambda_-)^{1/4} \\
 \pm i(\lambda_+\lambda_-)^{1/4}&\sqrt{\lambda_-} \end{array}\right).
\end{equation}
Components of the Stokes vectors of these states are $S_1^{(A)}=S_1^{(B)}=0$ and
\begin{equation}
 \label{St.-Schm-AB}
 \setlength{\extrarowheight}{0.1cm}
 \begin{array}{c}
 S_3^{(A)}=S_3^{(B)}=\displaystyle\frac{\sqrt{\lambda_+}-\sqrt{\lambda_-}}{\sqrt{1+C}}=\sqrt{\frac{1-C}{1+C}},\\
 S_2^{(A)}=-S_2^{(B)}=\displaystyle\frac{2(\lambda_+\lambda_-)^{1/4}}{\sqrt{1+C}}=\sqrt{\frac{2C}{1+C}}.
 \end{array}
\end{equation}
Thus, in the Schmidt-mode representation the Stokes vectors ${\vec S}^{(A)}$ and ${\vec S}^{(B)}$ are located in the vertical plane containing the horizontal axis $(+,-)$, and their directions are symmetric with respect to this axis. The angles between ${\vec S}^{(A,B)}$ and the biphoton Stokes vector ${\vec S}^{\,biph}\|(+,-)$ are determined by $S_3^{(A,B)}$-projections of the vectors ${\vec S}^{(A,B)}$ on the $(+,-)$:
\begin{equation}
 \label{SA-SB-half-angle}
 \cos\left(\theta_{AB}/2\right)=S_3^{(A)}=S_3^{(B)}=\sqrt{\frac{1-C}{1+C}}
 =\frac{P}{1+\sqrt{1-P^2}}=[A,B^\dag],
\end{equation}
where $\theta_{AB}$ is the angle between the vectors ${\vec S}^{(A)}$ and ${\vec S}^{(B)}$.

By inverting the relation between $\cos\left(\theta_{AB}/2\right)$ and the concurrence $C$, we can express the latter in terms of the angle $\theta_{AB}$ between the vectors ${\vec S}^{(A)}$ and ${\vec S}^{(B)}$:
\begin{equation}
 \label{C via theta-AB}
 C=\frac{1-\cos^2(\theta_{AB}/2)}{1+\cos^2(\theta_{AB}/2)}=\frac{1-\cos\theta_{AB}}{3+\cos\theta_{AB}}.
\end{equation}
And the expression for the degree of polarization $P$ via $\theta_{AB}$ has the form
\begin{equation}
 \label{P via theta-AB}
 P=\sqrt{1-C^2}=\frac{2\sqrt{2(1+\cos\theta_{AB})}}{3+\cos\theta_{AB}}=\frac{4|\cos(\theta_{AB}/2)|}{3+\cos\theta_{AB}}.
\end{equation}
This result for the degree of polarization agrees perfectly with that of the paper \cite{MCh}.

As the biphoton Stokes vector is directed along the bisector of the angle between the Stokes vectors ${\vec S}^{(A)}$ and ${\vec S}^{(B)}$, in the vectorial form the biphoton Stokes vector can be written as
\begin{eqnarray}
 \nonumber
 {\vec S}^{\,biph}=\left|{\vec S}^{\,biph}\right|\frac{{\vec S}^{(A)}+{\vec S}^{(B)}}
 {\left|{\vec S}^{(A)}+{\vec S}^{(B)}\right|}=P\frac{{\vec S}^{(A)}+{\vec S}^{(B)}}{2(1+\cos\theta_{AB})}\\
 \label{Vect S-bph}
 =\frac{2\cos(\theta_{AB}/2)}{(3+\cos\theta_{AB})(1+\cos\theta_{AB})}\left({\vec S}^{(A)}+{\vec S}^{(B)}\right).
\end{eqnarray}

Figure \ref{Fig.2}$b$ illustrates an alternative representation of a qutrit in terms  of the Schmidt-mode Stokes vectors ${\vec S}^{(+)}$ and ${\vec S}^{(-)}$.  The picture shows pairs of these Stokes vectors, oppositely directed and having the weight coefficients $\lambda_+$ and $\lambda_-$. The total biphoton Stokes vector equals the sum of all four Stokes vectors of the Schmidt modes with the weight coefficients, and half of this sum gives biphoton Stokes vector per one photon:
\begin{equation}
 \label{Sbf=S++S-}
 {\vec S}^{\,biph}=\lambda_+{\vec S}^{(+)}+\lambda_-{\vec S}^{(-)}\,.
\end{equation}

The results of this section are invariant with respect to rotations of the Poincar\'{e} sphere. The only changes concern the orientation of the $(+)-(-)$ axis and the plane containing the Stokes vectors ${\vec S}_A$ and ${\vec S}_B$. But at any orientation of the $(+)-(-)$ axis and $\left({\vec S}_A, {\vec S}_B\right)$-plane, the Stokes vectors of the Schmidt modes ${\vec S}^{(\pm)}$, of the factorizing operators ${\vec S}^{(A,B)}$, and of the biphoton state as a whole ${\vec S}^{\,biph}$ belong to the same plane. The biphoton Stokes vector is directed along the same direction as ${\vec S}^{(\pm)}$, and this direction coincides with the bisector of the angle $\theta_{AB}$ between ${\vec S}_A$ and ${\vec S}_B$. Eqs. (\ref{C via theta-AB})-(\ref{Sbf=S++S-}) remain valid without any changes in any frames with rotated Poincar\`{e} sphere. As to other derived equations (\ref{1ph-Schm-mode-st.vect}), (\ref{bf-stokes.v.-comp}), (\ref{S3=P}), (\ref{St.-Schm-AB}), and (\ref{SA-SB-half-angle}), they remain valid too, though the components of Stokes vectors $S_3$ and $S_2$ have to be understood as projections on the directions of the turned $(+)-(-)$ axis and turned $\left({\vec S}_A, {\vec S}_B\right)$-plane, rather than, e.g., projections on the $V-H$ and $R-L$ axes of the standard Poincar\'{e}-sphere orientation of Fig. \ref{Fig.1}.

\section{Possible experiments}
\begin{figure}[h]
\centering\includegraphics[width=12cm]{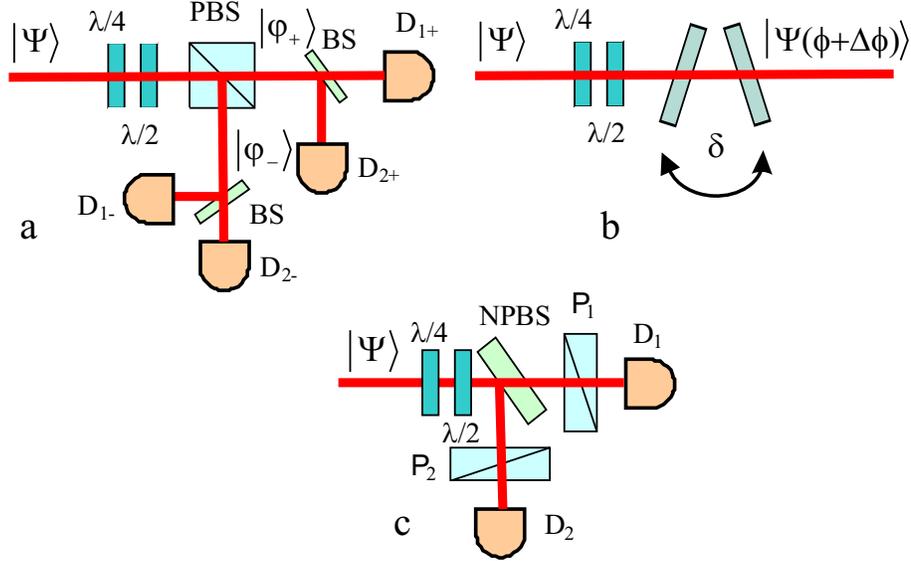}
\caption{{\protect\footnotesize {$a$: Experimental setup for observing polarization Schmidt modes of a qutrit $|\Psi\rangle$; $b$: A simple way to change the phase $\phi$ of the qutrit; $c$: Experimental setup for demonstrating polarization entanglement of a qutrit $|\Psi\rangle$.}}}\label{Fig.3}
\end{figure}

A possible experiment on the selection of polarization Schmidt modes of a qutrit is shown in Fig.~\ref{Fig.3}a. The qutrit state is sent to a polarization beam splitter (PBS) preceded by a quarter-wave and a half-wave plates. The plates are oriented in such a way that photons in the polarization Schmidt mode $|\varphi_+\rangle$ become horizontally polarized~\cite{plates}. Then the orthogonally polarized photons in the $|\varphi_-\rangle$ mode automatically become vertically polarized.
The qutrit state vector after the plates is transformed into a weighted superposition of photon pairs in vertical and horizontal modes,
\begin{equation}
|\Psi\rangle=\sqrt{\lambda_+}|2_H\rangle + e^{2i\phi}\sqrt{\lambda_-}|2_V\rangle,
\label{example}
\end{equation}
where $\phi$ is the relative phase of the Schmidt modes $|\varphi_-\rangle$ and $|\varphi_+\rangle$~\cite{phase}.

After the PBS, the pair $|2_H\rangle$ goes into the transmitted output port and the pair $|2_V\rangle$, into the reflected output port. Each photon pair can be detected as a coincidence of single-photon detector `clicks'. For registering such coincidences, beamsplitters (BS) followed by pairs of detectors are introduced in both output ports of the PBS. Then,  coincidences will be observed between the counts
of either detectors $D_{1+},D_{2+}$ or detectors $D_{1-},D_{2-}$, and their rates $R_\pm^c$ will scale
as the Schmidt eigenvalues $\lambda_{\pm}=R_{\pm}^c/(R_+^c+R_-^c)$. But no coincidences between the outputs of the detectors in different ports of the PBS will be observed. The polarization Schmidt modes can be found experimentally by finding positions of the quarter-wave and half-wave plates at which the coincidence counting rates between the detectors in  different output ports of the PBS turn to zero. This approach is most appropriate if the qutrit parameters $C_{1,2,3}$ are not known in advance. Further, from the experimentally found parameters $\lambda_{\pm}$, with the help of Eqs. (\ref{lambda}) one can easily calculate both the qutrit concurrence and degree of polarization.

In the absence of detectors' dark counts and parasite light, the beamsplitters in the PBS output ports are not necessary and only a single detector can be used in each port. Note that this method of finding $\lambda_{\pm}$ can be made invariant to the quantum efficiencies of the detectors. Indeed, instead of comparing the counting (coincidence) rates at the two output ports of the PBS one can compare the counting (coincidence) rates in the same port but at two different positions of the plates, corresponding to the Schmidt mode $|\varphi_+\rangle$ or $|\varphi_-\rangle$ sent to the same port.

It is worth mentioning that the four parameters determining a qutrit can be also introduced as the orientations of the plates transforming the qutrit from the form (\ref{Q-Schmidt}) into the one (\ref{example}), one of the Schmidt eigenvalues, and the phase $\phi$. Experimentally, the latter can be measured in the same scheme as shown in Fig. \ref{Fig.3}a but with the PBS rotated by $45^\circ$. Indeed, if the number of qutrits at the input of the PBS is $N$, and the quantum efficiencies of detectors $D_{1+}$ and $D_{2+}$ are, respectively, $\eta_1$ and $\eta_2$, then the numbers of their coincidences with the PBS oriented at $0^\circ$ and $90^\circ$ will be $R_{0}=\eta_1\eta_2\lambda_+N/2$ and $R_{90}=\eta_1\eta_2\lambda_-N/2$, respectively. At the same time, for a $45^\circ$ orientation of the PBS the number of coincidences will be $R_{45}=\eta_1\eta_2(1+2\sqrt{\lambda_+\lambda_-}\cos2\phi)N/4$. The visibility of the cosine-like dependence $R_{45}(\phi)$ is determined by the degree of entanglement and equals $C/(1+C)$. From the measurement of $R_{0}$, $R_{90}$ and $R_{45}$ one can infer both the Schmidt eigenvalues and the phase $\phi$ without knowing the quantum efficiencies and the initial number of pairs:
\begin{equation}
 \label{R-phi}
 \lambda_+=\frac{R_0}{R_0+R_{90}}, \lambda_-=\frac{R_{90}}{R_0+R_{90}}, \cos2\phi=\frac{2R_{45}-R_0-R_{90}}{2\sqrt{R_0R_{90}}}.
\end{equation}

Note that the phase $\phi$ in the expression (\ref{example}) is accessible for rather easy manipulations. Indeed, by providing some delay time $\Delta t$ for the vertically polarized photons after PBS in the scheme of Fig. \ref{Fig.3}$a$ and then combining photons from both channels into a single beam, one gets the qutrit state of the same form as in Eq. (\ref{example}) but with a different phase $\phi\rightarrow\phi+\Delta\phi$, where $\Delta\phi=\omega_{ph}\Delta t$ and $\omega_{ph}$ is the photon frequency. If one makes $\Delta\phi=\pi-\phi$, the phase factor in front of $|2_V\rangle$ in Eq. (\ref{example}) becomes equal to $e^{2\pi\,i}=1$, and the qutrit state vector takes the form identical to that of Eq. (\ref{Decomp-st-vect}) but with the operators $a_+^\dag$ and $a_-^\dag$ substituted by $a_H^\dag$ and $a_V^\dag$: $|\Psi\rangle\rightarrow\frac{1}{\sqrt{2}}(\sqrt{\lambda_+}a_H^{\dag^2} +\sqrt{\lambda_-}a_V^{\dag^2}) |0\rangle$. A structure of the Stokes vectors of the qutrit reduced to this form is identical to that shown in Fig. \ref{Fig.2}a but with the vector $\vec{S}^{biph}$ oriented along the $H$ axis. This shows that the same structure of Stokes vectors  as in Fig. \ref{Fig.2} can be obtained in two ways: either by appropriate rotations of the Poincar\'{e} sphere for a given arbitrary qutrit or by the above described transformation of the qutrit itself with the help of the $\lambda/4$- and $\lambda/2$-plates plus the time delay in one of the channels in Fig. \ref{Fig.3}$a$ canceling the phase $\phi$ in Eq. (\ref{example}), but with the orientation of the Poincar\'{e} sphere  kept standard (the `active' and `passive' viewpoints discussed in Ref.~\cite{Klyshko97}).

In practice, the phase delay $\Delta\phi$ is most conveniently introduced in a polarization interferometer as shown in Fig.~\ref{Fig.3}b. After the qutrit is transformed to the form (\ref{example}) by means of two waveplates, it is transmitted through two birefringent plates of thickness $l$ with the optic axes horizontal. The plates are tilted symmetrically so that the angle between them is $\delta$. Due to the birefringence $\Delta n$, a vertically polarized photon pair in Eq.~(\ref{example}) acquires a phase $\Delta\phi=2\Delta n l/\cos(\delta/2)$ with respect to the horizontally polarized pair.

This simple way to modify the state is only accessible for an entangled qutrit. For a disentangled (fully polarized) qutrit, the phase $\phi$ has no meaning, and the only way to transform the state is to subject it to some polarization transformation, the same way it can be done with a qubit. This provides the encoding of only two numbers. Thus, an entangled qutrit can undergo a more general transformation, characterized by three parameters, while the transformation of a disentangled qutrit is only described by two parameters. This shows unambiguously that entangled biphoton qutrits possess additional operational properties, which can be used directly in applications without transition to higher-dimensionality states.

For further physical interpretation of the polarization entanglement of a qutrit, one can suggest another experimental scheme (Fig.~\ref{Fig.3}c). The qutrit $|\Psi\rangle$, after the polarization transformation, is directed to a $50\%$ non-polarizing beam splitter (NPBS). Half of the photon pairs will exit into a single output port of the NPBS, but the other half will be split between the two ports. This part of the state can be written as

\begin{equation}
 \label{working_class}
|\Psi\rangle_{12}\equiv \sqrt{\lambda_+}\;|1_{H1},1_{H2}\rangle  +e^{2i\phi}\sqrt{\lambda_-}\;|1_{V1},1_{V2}\rangle,
\end{equation}
\noindent where the indices $H1$ ($H2$) denote horizontal polarization in the transmitted (reflected) beam, and similarly for the vertical polarization. Note that the photon pairs exiting into a single NPBS port will not affect the coincidence counting rate for detectors $D_1$ and $D_2$. One can say that the state (\ref{working_class}) is postselected by counting such coincidences.


This state has been considered in Ref.~\cite{Kwiat}, and it was shown to be maximally entangled  only if the Schmidt eigenvalues are equal, $\lambda_+=\lambda_-$. This exactly corresponds to the case where $C=1$ and the qutrit $|\Psi\rangle$ is fully entangled. If $\lambda_+\ne\lambda_-$, the state is non-maximally entangled, and its degree of entanglement can be measured, as shown in Ref.~\cite{Kwiat}, from the orientations of the polarizers $P_1,P_2$ at which the coincidence counting rate for detectors $D_1,D_2$ vanishes.

It is also worth mentioning that (\ref{working_class}) represents the state into which a qutrit should be transformed to be applicable for quantum teleportation~\cite{SPK-Mol}. The degree of entanglement of this state and, hence, its applicability for quantum teleportation is determined by the degree of entanglement of the initial qutrit.

\section{Conclusion}
We have considered the general case of a biphoton qutrit, physically represented by a photon pair in a single frequency and wavevector mode. We have suggested a simple procedure for finding explicitly and in a general form expressions for the single-photon creation operators $A^\dag$ and $B^\dag$, factorizing the qutrit state vectors, i.e., representing them in the form  $A^\dag B^\dag|0\rangle$. We have shown that the described operator factorization of state vectors is not related to the factorization of the biphoton polarization wave function and does not exclude a possibility of entanglement of biphoton qutrits. The degrees of their entanglement and polarization are found to be related directly to the commutation features of the factorizing operators by Eqs. (\ref{Comm-Schm}) and (\ref{C via comm}). Qutrits are shown to be disentangled only if the commutator $[A,B^\dag]$ equals unity, i.e., if the factorizing creation operators coincide with each other, $A^\dag=B^\dag$. This means that all disentangled  biphoton qutrits are representable in the form $\frac{1}{\sqrt{2}}A^{\dag\,^2}|0\rangle$ with an arbitrary single-photon creation operator $A^\dag$. In all other cases, when $[A,B^\dag]\neq 1$ and, hence, $A^\dag\neq B^\dag$, biphoton qutrits are entangled. Qutrits are maximally entangled (and unpolarized) if $[A,B^\dag]=0$. In this case the operators $A^\dag$ and $B^\dag$ characterize orthogonal polarization modes, and a typical example is the state $a_H^\dag a_V^\dag|0\rangle$. Further, we have considered the polarization Schmidt decomposition of a generic biphoton qutrit.We have found direct relations between the factorizing operators $A^\dag$, $B^\dag$ and the Schmidt-mode creation operators $a_+^\dag,\,a_-^\dag$ (Eqs. (\ref{AB-SCHM}),\,(\ref{a+,a- via AB})). We have suggested a new image of biphoton qutrits on the Poincar\'{e} sphere in terms of the Stokes vectors of the Schmidt modes. The structure of all Stokes vectors characterizing biphoton qutrits takes the simplest form in the Schmidt-mode representation, which corresponds to a very specific orientation of the Poincar\'{e} sphere shown in Fig. \ref{Fig.2}. The simplest form of the Schmidt decomposition for biphoton qutrits is shown to be given by Eq.  (\ref{example}) with the sum of squared creation operators of horizontally and vertically polarized photons and a phase factor between them. A procedure is proposed for measuring  both the degree of entanglement of this state and the phase factor. A way of simple manipulations with the phase is described and is suggested to be used for the encoding  of quantum information. A physical interpretation is proposed for the polarization entanglement of a biphoton qutrit.

\ack{This work was supported in part by the Russian Foundation for Basic Research, grant \#11-02-01074.}


\begin{thebibliography}{13}
\bibitem{phase} Despite the fact that the polarization state of a single photon is usually considered with the total phase meaningless and therefore set equal to zero, the relative phase $\phi$ between the Schmidt modes $\varphi_-$ and $\varphi_+$ is important and cannot be omitted.
\bibitem{plates} This is always possible as such a combination of plates suffices for transforming any polarization state of light into the linear polarization state transmitted by the PBS.
\end{thebibliography}
\end{document}